\documentclass[12pt,psfig]{article}
\usepackage{amssymb}
\usepackage{amsmath,amsthm}
\usepackage{amsfonts}
\usepackage{graphicx}
\usepackage{graphics}
\numberwithin{equation}{section}

\usepackage{hyperref}

\begin{document}
\begin{center}\Large\textbf{Effective Action of a Dynamical
D$p$-brane with Background Fluxes}
\end{center}
\vspace{0.75cm}
\begin{center}{\large Shiva Heidarian and \large Davoud
Kamani}
\end{center}
\begin{center}
\textsl{\small{Physics Department, Amirkabir University of
Technology (Tehran Polytechnic)\\
P.O.Box: 15875-4413, Tehran, Iran\\
e-mails: kamani@aut.ac.ir ,
sh.Heidarian@aut.ac.ir\\}}
\end{center}
\vspace{0.5cm}

\begin{abstract}

We shall construct the Dirac-Born-Infeld like and
the Wess-Zumino like actions for
a dynamical D$p$-brane with the
$U(1)$ gauge potential and the Kalb-Ramond
background field.
The brane dynamics simultaneously has both tangential
and transverse components.
Our calculations will be in the context
of the type II superstring theory,
via the boundary state formalism.

\end{abstract}


\newpage
\section{Introduction}

Since D-branes could be defined as hypersurfaces on which
the boundaries of the string worldsheets can end on them,
the boundary state formalism elaborates
a perfect description of them
\cite{1}-\cite{10}.
This method obviously provides a closed string
description of the D-branes, and is applicable
to various CFTs
that have been used for different configurations
of the D-branes, e.g. see Refs. \cite{2},
\cite{5}-\cite{11} and references therein.

On the other hand, dynamics of a D-brane is
properly described in the field theory
by an effective action which is given by the sum of
the Dirac-Born-Infeld (DBI) \cite{12}-\cite{18} and the
Wess-Zumino (WZ) \cite{3}, \cite{19}-\cite{21} actions.
These are low energy effective actions
of the massless fields which are induced on the brane
worldvolume. In other words,
these actions prominently specify the interactions between
the D-brane and the massless fields.
However, one of the importance of these actions is that
they are indicative of the various dualities
of the string theories \cite{22}-\cite{25}.

A D-brane action can be extracted by various rules:
the string $\sigma$-model approach
\cite{17}, \cite{18}, \cite{26},
the background independent open string field theory
\cite{13}, \cite{27},
the scattering amplitude approach \cite{28}, \cite{29}, and
the boundary state method \cite{2}, \cite{3},
\cite{25}, \cite{30}-\cite{34}.
A D-brane couples to the graviton, Dilaton
and Kalb-Ramond fields,
and since it carries the Ramond-Ramond charges \cite{1},
it couples to the R-R fields too.
Therefore, a boundary state, which reveals
the couplings of the closed string states with the D-brane,
gives us a satisfactory method to calculate
the D-brane action and its corrections via
the boundary actions of the fundamental string.

In this paper we shall obtain
the DBI-like and WZ-like actions for
a single dynamical D$p$-brane
with background fields. Our method
is the boundary state formalism in the framework of
the type IIA and type IIB superstring theories.
The  brane has been
dressed by a $U(1)$ gauge potential $A_{\alpha}$ and
a constant antisymmetric field $B_{\mu\nu}$.
Besides, the brane has a uniform rotation inside its
volume and a uniform linear motion with both
transverse and tangential components.
We shall see that the background fields and dynamics
of the brane extremely influence the boundary
state of the brane, and hence the resultant action.
This generalized effective action of the brane, due
to the fundamental role of the D-branes,
is presumably valuable and may possibly give
a deeper understanding of the substantial properties
of the D-branes.

This paper is organized as follows.
In Sec. 2, we shall introduce the
boundary states of the NS-NS and R-R sectors
of superstring, corresponding to a
dressed-dynamical D$p$-brane.
In Sec. 3, a DBI-like action for this brane
will be constructed.
In Sec. 4, a WZ-like action for the same brane
will be built.
Section 5 is devoted to the conclusions.

\section{The boundary state of a dressed-dynamical D-brane}
\hspace{0.5cm}

A boundary state is a closed string state
that manifestly encodes
all properties of the corresponding D-brane such as:
the brane couplings with the closed string states,
the brane tension, dynamical variables of the brane
and internal fields. This adequate state clarifies that a
D-brane can emit (absorb) all closed string states.
Therefore, a D-brane can be completely described
by an appropriate boundary state.

We begin with the following sigma-model action
for closed string to compute the boundary state,
associated with a dressed-dynamical
D$p$-brane,
\begin{eqnarray}
S_{\rm bulk}&=&-\frac{1}{4\pi \alpha'}\int_{\Sigma}d^2\sigma
\left(\sqrt{-g}g^{ab}G_{\mu\nu}\partial_{a}X^{\mu}
\partial_b X^{\nu}+\epsilon^{ab}B_{\mu\nu}\partial_a X^{\mu}
\partial_b X^{\nu}\right),
\nonumber\\
S_{\rm bdry}&=&\frac{1}{2\pi \alpha'}
\int_{\partial\Sigma}d\sigma
\left(A_{\alpha}\partial_{\sigma}X^{\alpha}+
\omega_{\alpha\beta}J^{\alpha\beta}_{\tau}\right),
\end{eqnarray}
where the total action is
$S=S_{\rm bulk}+S_{\rm bdry}$.
The coordinates $\{x^{\alpha}|\alpha=0,1,\ldots,p\}$
indicate the directions along the brane worldvolume.
We apply a constant Kalb-Ramond field $B_{\mu\nu}$
and the well-known gauge
$A_{\alpha}=-\frac{1}{2}F_{\alpha \beta }X^{\beta}$
with a constant field strength $F_{\alpha \beta }$
for the $U(1)$ gauge potential.
The string worldsheet and background spacetime
are flat with
$G_{\mu\nu}={\rm diag} (-1,1,\ldots,1)$.
The tangential dynamics of the brane
consists of a constant antisymmetric angular
velocity $\omega_{\alpha\beta}$
which represents the tangential linear motion
and rotation of the brane,
and $J^{\alpha \beta}_\tau=X^\alpha
\partial_\tau X^\beta-X^\beta \partial_\tau X^\alpha$
shows the angular momentum density.
The parameters $\omega_{{\bar \alpha}{\bar \beta}}$
and $\omega_{0{\bar \alpha}}$, with
${\bar \alpha},{\bar \beta}\in \{1,2,\ldots,p\}$,
denote the angular and linear
velocities of the brane, respectively.
A transverse linear motion will be also added to the brane.
Note that presence of the background fields
specifies some preferred alignments inside
the brane worldvolume. Thus,
the Lorentz symmetry in the worldvolume subspace
has been explicitly broken. This elucidates that
the tangential dynamics of the brane is meaningful.

Now we impose a transverse velocity to the brane.
At first, vanishing the variation of the action (2.1)
defines the primary boundary state equations.
Then, we introduce a Lorentz boost
along the transverse direction $x^{i_0}$
with the velocity $v^{i_0} \equiv v$ into the foregoing
boundary state equations. Hence,
the boosted boundary state equations
possess the features
\begin{eqnarray}
&~&[\partial_{\tau}(X^0-v X^{i_0})
+4\omega^{0}_{~~{\bar\beta}}
\partial_\tau X^{{\bar \beta}}
+\mathcal{F}^{0}_{~~{\bar\beta}}
\partial_\sigma X^{{\bar \beta}}]_{\tau=0}
|B_x\rangle=0~,
\nonumber\\
&~&[\partial_{\tau}X^{\bar{\alpha}}
+4\gamma^2\omega^{\bar{\alpha}}_{~~0}
\partial_\tau (X^0-v X^{i_0})
+4\omega^{\bar{\alpha}}_{~~{\bar \beta}}
\partial_\tau X^{\bar \beta}
\nonumber\\
&~&+\gamma^2\mathcal{F}^{\bar{\alpha}}_{~~0}
\partial_\sigma (X^0-v X^{i_0})
+\mathcal{F}^{\bar{\alpha}}_{~~{\bar \beta}}
\partial_\sigma X^{\bar \beta}]_{\tau=0}
|B_x\rangle=0~,
\nonumber\\
&~&(X^{i_0}-v X^{0}-y^{i_0})_{\tau=0}
|B_x\rangle=0~,
\nonumber\\
&~&(X^i-y^i)_{\tau=0}|B_x\rangle=0~,
\end{eqnarray}
where $\gamma =1/\sqrt{1-v^2}$,
$i\in \{p+1,\ldots,{\hat {i_0}}, \ldots,9\}$, i.e.
$i \neq {i_0}$,
the set $\{y^i, y^{i_0}\}$ indicates
the initial location of the brane, and
$\mathcal{F}_{\alpha\beta}=B_{\alpha\beta}-F_{\alpha\beta}$
is the total field strength.

Introducing the mode
expansion of the closed string coordinates
$X^\mu (\sigma , \tau)$
into Eqs. (2.2) gives these equations in
terms of the string oscillators and
zero-modes $\alpha^\mu_n$, ${\tilde \alpha}^\mu_n$,
$x^\mu$ and $p^\mu$. The resultant
equations can be solved by the coherent state
method to produce the boundary state,
\begin{eqnarray}
|B_x\rangle&=&\frac{T_p}{2}
\sqrt{-\det Q}\;
\exp\left[{-\sum_{m=1}^{\infty}
\left(\frac{1}{m}
\alpha_{-m}^{\mu}S_{\mu\nu}
\tilde{\alpha}_{-m}^{\nu}\right)}\right]
|0\rangle_\alpha
\otimes|0\rangle_{\tilde{\alpha}}~\label{aos}
\nonumber\\
&\times&~\delta\left({x}^{{i_0}}
-vx^{0}-y^{{i_0}}\right)|p^{i_0}=0\rangle
 \prod_i\left[ \delta \left({x}^i
-y^i\right)|p^i=0\rangle
\right]\prod_{\alpha}|p^{\alpha}=0\rangle,
\end{eqnarray}
where $T_p=\sqrt{\pi}(4\pi^2\alpha')^{(3-p)/2}$
is related to the D$p$-brane tension, and
the matrices have the following definitions
\begin{eqnarray}
S_{\mu\nu}&=&\left((Q^{-1}N)_{\lambda\lambda'},
-\delta_{ij}\right)~,
\nonumber\\
Q_{~~\lambda}^{0}&=&
\gamma(\delta^{0}_{~~\lambda}
-v\delta^{{i_0}}_{~~\lambda})
+\gamma(4\omega^{0}_{~~{\bar\alpha}}
-\mathcal{F}^{0}_{~~\bar{\alpha}})
\delta^{\bar{\alpha}}_{~~\lambda}~,
\nonumber\\
Q_{~~\lambda}^{\bar{\alpha}}&=&
\delta^{\bar{\alpha}}_{~~\lambda}+
\gamma^{2}(4\omega^{{\bar\alpha}}_{~~0}
-\mathcal{F}^{\bar{\alpha}}_{~~0})
(\delta^{0}_{~~\lambda}
-v\delta^{{i_0}}_{~~\lambda})+
(4\omega^{{\bar\alpha}}_{~~\bar{\beta}}
-\mathcal{F}^{\bar{\alpha}}_{~~\bar{\beta}})
\delta^{\bar{\beta}}_{~~\lambda}~,
\nonumber\\
Q_{~~\lambda}^{{i_0}}&=&\delta^{{i_0}}_{~~\lambda}
-v\delta^{0}_{~~\lambda}~,
\nonumber\\
\nonumber\\
N_{~~\lambda}^{0}&=&
\gamma(\delta^{0}_{~~\lambda}
-v\delta^{{i_0}}_{~~\lambda})
+\gamma(4\omega^{0}_{~~{\bar\alpha}}
+\mathcal{F}^{0}_{~~\bar{\alpha}})
\delta^{\bar{\alpha}}_{~~\lambda}~,
\nonumber\\
N_{~~\lambda}^{\bar{\alpha}}&=&
\delta^{\bar{\alpha}}_{~~\lambda}+
\gamma^{2}(4\omega^{{\bar\alpha}}_{~~0}
+\mathcal{F}^{\bar{\alpha}}_{~~0})
(\delta^{0}_{~~\lambda}-
v\delta^{{i_0}}_{~~\lambda})+
(4\omega^{{\bar\alpha}}_{~~\bar{\beta}}
+\mathcal{F}^{\bar{\alpha}}_{~~\bar{\beta}})
\delta^{\bar{\beta}}_{~~\lambda}~,
\nonumber\\
N_{~~\lambda}^{{i_0}}&=&-\delta^{{i_0}}_{~~\lambda}
+v\delta^{0}_{~~\lambda}~,
\end{eqnarray}
where $\lambda , \lambda' \in \{0,1, \ldots,p~;i_0\}$.

Since we shall utilize the overall factor of
Eq. (2.3), let briefly describe its derivation.
The convenient gauge
$A_\alpha = -\frac{1}{2}F_{\alpha\beta}X^\beta$,
accompanied by the quadratic tangential dynamics term,
impose a quadratic form to the boundary action
of Eq. (2.1). Path integration on this Gaussian action
obviously produces the prefactor
$\prod_{n=1}^\infty \left(-\det Q\right)^{-1}$
in Eq. (2.3). Applying the regularization scheme
$\prod_{n=1}^\infty a^{-1} \longrightarrow \sqrt{a}$
recasts the normalization factor to
$\sqrt{-\det Q}$. For such overall factors of
the stationary setups e.g. see
the Refs. \cite{12}, \cite{13}, \cite{35}.

In fact, the coherent state method
provides the boundary state (2.3)
under the condition $SS^{T}=\mathbf{1}$.
This condition introduces the following
relations among the input parameters
\begin{eqnarray}
&~& \omega^{0}_{~~\bar{\alpha}}
~\mathcal{F}^{0}_{~~\bar{\alpha}}=0,
\nonumber\\
&~&\omega^{0 \bar{\beta}}
~\mathcal{F}^{\bar{\alpha}}_{~~\bar{\beta}}
+\mathcal{F}^{0 \bar{\beta}}
~\omega^{\bar{\alpha}}_{~~\bar{\beta}}=0,
\nonumber\\
&~&\mathcal{F}^{\bar{\alpha}}_{~~\bar{\kappa}}
~\omega^{\bar{\beta} \bar{\kappa}}+
\mathcal{F}^{\bar{\beta}}_{~~\bar{\kappa}}
~\omega^{\bar{\alpha} \bar{\kappa}}-
\gamma^{2}\left(\omega^{\bar{\alpha}}_{~~0}
~\mathcal{F}^{\bar{\beta}}_{~~0}+
\omega^{\bar{\beta}}_{~~0}
~\mathcal{F}^{\bar{\alpha}}_{~~0}\right)=0~.
\end{eqnarray}
Thus, from the total $1+3p(p+1)/2$ parameters of the set
$\{\omega_{\alpha\beta},F_{\alpha\beta},B_{\alpha\beta} ,v\}$
only $p^2$ of them are independent.

The worldsheet supersymmetry guides us
to employ the
following replacements on the bosonic boundary state
equations (2.2) to
construct conveniently their fermionic counterparts
\begin{eqnarray}
\partial_{+}X^\mu(\sigma ,\tau)&\rightarrow&
-i\eta\psi_{+}^\mu(\tau +\sigma)~,
\nonumber\\
\partial_{-}X^\mu(\sigma,\tau)&\rightarrow&
-\psi_{-}^\mu(\tau - \sigma)~,
\label{jk}
\end{eqnarray}
where $\partial_\pm = (\partial_\tau \pm \partial_\sigma)/2$,
and $\eta=\pm1$ will be used for the GSO projection.
Therefore, the boundary state equations
of the worldsheet fermions,
in terms of the fermionic oscillators,
find the following features
\begin{eqnarray}
&~& \left(\psi_t^\mu-i\eta S^\mu_{~~\nu}
\tilde{\psi}^{\nu}_{-t}\right)
|B_\psi^{\rm (osc)};\eta\rangle_{\rm R, NS}=0~,
\nonumber\\
&~& \left(\psi_0^\mu-i\eta S^\mu_{\;\;\;\nu}
\tilde{\psi}^{\nu}_{0}\right){|B_\psi^{(0)};
\eta\rangle_{\rm R}}=0~,
\end{eqnarray}
where the decomposition
$|B_{\psi};\eta\rangle=|B^{\rm (osc)}_{\psi};\eta\rangle
\otimes|B^{(0)}_{\psi};\eta\rangle$ was applied, and
$t\in \mathbb{Z}-\{0\}\;(t\in \mathbb{Z}+1/2)$ is related
to the R-R (NS-NS) sector.

Solutions of Eqs. (2.7) are given by
\begin{eqnarray}
|B_\psi;\eta\rangle_{\rm NS}
=-i\exp\left[i\eta\sum_{r=1/2}^{\infty}\psi_{-r}^\mu~
S_{\mu\nu}~\tilde{\psi}_{-r}^{\nu}\right]
{|0\rangle_{\rm NS}}~,
\end{eqnarray}
\begin{eqnarray}
|B_\psi;\eta\rangle_{\rm R}
&=& -\frac{\gamma}{\sqrt{-\det Q}}
~\exp\left[i\eta\sum_{n=1}^{\infty}\psi_{-n}^\mu~
S_{\mu\nu}~\tilde{\psi}_{-n}^{\nu}\right]
\nonumber\\
&\times&\left({C}(\Gamma^{0}
+v\Gamma^{{i_0}})\;\Gamma^1\ldots\Gamma^p~
\frac{1+i\eta{\Gamma^{11}}}{1+i\eta}~\mathcal{H}\right)_{AB}~
|A\rangle \otimes|\tilde{B}\rangle~,
\nonumber\\
\mathcal{H} &=& \left[ 1+v\Gamma^{i_0}\Gamma^{0} -2
v\Gamma^{{i_0}}\Gamma^{0}\left(1
+\left(PQ^{-1}N\right)^{i_0}_{~~\lambda''}\Gamma^{i_0}
\Gamma^{\lambda''}\right)^{-1}\right]^{-1}
\nonumber\\
&\times& :\exp\left(-\frac{1}{2}{{\Phi}}_{\lambda\lambda'}
\Gamma^\lambda\Gamma^{\lambda'}\right):~,
\nonumber\\
{\Phi}&=&(\phi-\phi^T)/2~,~~~~~~~
\phi_{\lambda\lambda'} \equiv
\left((PQ^{-1}N+1)^{-1}(PQ^{-1}N-1)\right) _{\lambda\lambda'}~,
\end{eqnarray}
where $C$ is the matrix of charge conjugation, and
$|A\rangle$ and $|\tilde{B}\rangle$ are spinor vacua.
The matrix $P$ is defined by
$P^{\lambda}_{~~\lambda'}=(\delta^{\alpha}
_{~~\beta}\;,-\delta^{i_0}_{~~{i_0}})$ with
$P_{\alpha i_0}=P_{i_0 \alpha}=0$.
The conventional notation $:~:$ implies that
we should expand the exponential factor
with the convention that all Dirac matrices anticommute,
thus, a finite number of terms remain.
If the dynamical variables vanish we receive
$\Phi_{\alpha\beta}=\mathcal{F}_{\alpha\beta}$
which is consistent with the
results of the literature.

For eliminating the tachyonic state and
preserving the supersymmetry
the GSO projection should be applied.
The total boundary state of each sector,
after the GSO projection,
is given by a linear combination of the
boundary states with $\eta =\pm 1$,
\begin{eqnarray}
&~&|B\rangle_{\rm NS}=\frac{1}{2}
\left(|B;+\rangle_{\rm NS}
-|B;-\rangle_{\rm NS}\right)~,
\nonumber\\
&~&|B\rangle_{\rm R}=\frac{1}{2}
\left(|B;+\rangle_{\rm R}
+|B;-\rangle_{\rm R}\right)~,
\nonumber\\
&~&|B;\eta\rangle_{\rm NS,R}=
|B_x\rangle\otimes
|B_\psi;\eta\rangle_{\rm NS,R}
\otimes|B_{\rm gh}\rangle
\otimes|B_{\rm sgh};\eta\rangle_{\rm NS,R}~,
\end{eqnarray}
where $|B_{\rm gh}\rangle$ and
$|B_{\rm sgh}\rangle$ are the known boundary
states corresponding to the conformal
and superconformal ghosts, respectively.
They are independent of the background fields
and the brane dynamics.

In the next two sections we shall utilize the
GSO-projected boundary states to extend the
action of a stationary D$p$-brane, i.e.
$S_{{\rm D}p}=S_{\rm DBI}+S_{\rm WZ}$,
to our dynamical-dressed D$p$-brane.
Note that for each setup the corresponding D-brane
action accurately represents the interactions
of the brane with the massless fields.

\section{The DBI-like action}

In one hand we have the DBI action which
reveals the couplings of
the brane with the graviton, dilaton and
Kalb-Ramond fields. On the other hand,
since the boundary state encodes all properties
of the brane, it also reproduces the same
couplings between the brane and the massless states of
closed string. Meanwhile, the disk partition function
\cite{12}, \cite{13}, \cite{35},
which is proportional to the inner product
$\langle {\rm vacuum}| B \rangle_{\rm NS}$,
elucidates that the normalization
factor of the NS-NS boundary state
nearly defines a DBI-like Lagrangian.
Thus, the DBI-like action for our brane is given by
\begin{eqnarray}
S^{(\omega , v)}_{\rm DBI}=-\frac{T_p}{\kappa}
\int d^{p+1} \xi \sqrt{-\det {\tilde Q}_{\lambda \lambda'}}~,
\end{eqnarray}
where $\kappa = (2\pi)^{7/2}(\alpha')^2 g_s/\sqrt{2}$ is
the gravitational constant and $g_s$ is the string coupling.
The matrix ${\tilde Q}_{\lambda \lambda'}$
is closely related to $Q^\lambda_{~~\lambda'}$, i.e.,
we should apply the pull-back of the metric and
Kalb-Ramond field to the matrix $Q$ to obtain ${\tilde Q}$.
However, this is a generalized DBI action
which is corresponding to a dynamical D$p$-brane
with background fields. The explicit form of the
matrix $Q$, i.e. Eq. (2.4), shows the combination
$\left( 4\omega-\mathcal{F}\right)_{\alpha \beta}$
in the action (3.1). This clarifies that
the tangential dynamics and the internal parts of the
background fields appear in a similar fashion
in the DBI-like action.
However, for the stationary branes, i.e. by quenching
$\omega$ and $v$, the DBI-like action (3.1) reduces to
the conventional DBI action, as expected.

As a special case, by stopping the transverse
motion we obtain the action
\begin{eqnarray}
S^{(\omega , 0)}_{\rm DBI}&=&
-\frac{T_p}{\kappa}\int d^{p+1} \xi
\sqrt{-\det \left[{\tilde G}_{\alpha\beta}
+{\tilde B}_{\alpha\beta}-2\pi\alpha'(F_{\alpha\beta}
-4\omega_{\alpha\beta})\right]}~,
\end{eqnarray}
where ${\tilde G}_{\alpha\beta}$ and
${\tilde B}_{\alpha\beta}$ are pull-back
of $G_{\mu\nu}$ and $B_{\mu\nu}$ on the
brane worldvolume, respectively.
In fact, we applied the static gauge $\xi^{i_0}=X^{i_0}$
which simplified the elements of the induced metric
${\tilde G}_{\lambda \lambda'}$
as ${\tilde G}_{i_0 i_0}=1$, ${\tilde G}_{\alpha i_0}=0$
and ${\tilde G}_{\alpha\beta}$.
Expansion of this action for
$\omega_{\alpha\beta}<<1$ yields
the usual DBI action and its corrections
due to the tangential dynamics.
Therefore, by using the formula
\begin{eqnarray}
\sqrt{\det (M_0+M)}&=&\sqrt{\det M_0}\bigg{[}
1+\frac{1}{2}{\rm Tr}\left( M_0^{-1}M \right)
\nonumber\\
&-&\frac{1}{4}{\rm Tr}\left( M_0^{-1}M \right)^2
+\frac{1}{4}\left[{\rm Tr}\left(M_0^{-1}M
\right)\right]^{2}+\ldots \bigg{]},
\end{eqnarray}
we acquire the following $\omega$-corrections
\begin{eqnarray}
S^{(\omega ,0)}_{\rm DBI}&=&-\frac{T_p}{\kappa}\int d^{p+1} \xi
\sqrt{-\det \left({\tilde G}_{\alpha\beta}
+{\tilde B}_{\alpha\beta}-2\pi\alpha'F_{\alpha\beta}\right)}
\nonumber\\
&\times &\bigg{\{} 1+4\pi{\alpha'}{\rm Tr}
\left[({\tilde G}+2\pi\alpha'\mathcal{\bar F})^{-1}\omega\right]
-16\pi^2{\alpha'}^2 {\rm Tr}
\left[({\tilde G}+2\pi\alpha'\mathcal{\bar F})^{-1}\omega\right]^2
\nonumber\\
&+& 16\pi^2{\alpha'}^2 \left({\rm Tr}
\left[({\tilde G}+2\pi\alpha'
\mathcal{\bar F})^{-1}\omega\right]\right)^2
+\ldots \bigg{\}},
\end{eqnarray}
where $2\pi\alpha'\mathcal{\bar F}={\tilde B}_{\alpha\beta}
-2\pi\alpha'F_{\alpha\beta}$.

As another special case, quench the tangential
dynamics. In this case working with an
arbitrary D$p$-brane does not give the explicit
form of the action. Hence, we consider a
D3-brane with the velocity $v$ along the $x^4$-direction.
Furthermore, let the $4\times 4$ antisymmetric matrix
$\mathcal{F}_{\alpha\beta}$ be skew-diagonal,
i.e. block-diagonal with two $2\times 2$ antisymmetric
matrices. The nonzero elements of the blocks
are $2\pi\alpha'f$, $-2\pi\alpha'f$, $2\pi\alpha'g$
and $-2\pi\alpha'g$.
Thus, the matrix ${\tilde Q}_{~\lambda'}^{\lambda}$
possesses the structure
\begin{eqnarray}
{\tilde Q}_{~\lambda'}^{\lambda}=\left(
\begin{array}{ccccc}
\gamma &-2\pi\alpha'\gamma f &0&0& -v\gamma\\
-2\pi\alpha'{\gamma}^{2}f &1 &0&0 & 2\pi\alpha'{\gamma}^{2}vf\\
0&0&1&-2\pi\alpha'g & 0 \\
0& 0& 2\pi\alpha'g &0 &0 \\
-\gamma v& 0&0&0 &\gamma \\
\end{array}\right).
\end{eqnarray}
For constructing the action we apply
${\tilde Q}_{\lambda \lambda'}={\tilde G}_{\lambda \lambda''}
{\tilde Q}^{\lambda''}_{~ \lambda'}$ and the static gauge
$\xi^{i_0}=X^{i_0}$.
In this case the generalized action (3.1) reduces to
\begin{eqnarray}
S^{(0,v)}_{\rm DBI}=
-\frac{\gamma T_3}{\kappa}\int d^{4} \xi
\sqrt{-\det \left({\tilde G}_{\alpha\beta}
+{\tilde B}_{\alpha\beta}-2\pi\alpha'F_{\alpha\beta}\right)
+v^2\left(1+(2\pi\alpha'g)^2\right)
\det {\tilde G}_{\alpha\beta}}~.
\end{eqnarray}

For a very small speed, i.e. $v<<1$,
the action (3.6) is decomposed into
the conventional DBI action and its velocity corrections
\begin{eqnarray}
S^{(0,v)}_{\rm DBI}&=&-\frac{T_3}{\kappa}\int d^4 \xi
\sqrt{-\det \left({\tilde G}_{\alpha\beta}
+{\tilde B}_{\alpha\beta}-2\pi\alpha'F_{\alpha\beta}
\right)}
\nonumber\\
&\times& \left[1 +\frac{v^2}{2}\left(
1-\frac{1+(2\pi\alpha'g)^2}
{\det(1+2\pi \alpha'{\tilde G}^{-1}
\mathcal{\bar F})}\right) +\ldots \right].
\end{eqnarray}
We observe that the lowest order correction
is the second order of $v$, while Eq. (3.4) demonstrates that
the corrections due to the tangential dynamics
begin with the first order of $\omega$.
The generalization (3.6) and the velocity correction
(3.7) are compatible with the literature,
specially for the D0-branes, e.g. see \cite{15, 36, 37}.
\subsection{Some note on the DBI-like action}

\subsubsection{The Yang-Mills theory}

It is well known that the $(p+1)$-dimensional Yang-Mills theory
can be extracted from the D$p$-brane effective action.
For example, Eq. (3.7) illustrates that the Yang-Mills
theory, which lives on the
worldvolume of the D3-brane with a slow transverse
motion, possesses the coupling constant
$g_{\rm YM}=\frac{\sqrt{\kappa}}{\pi\alpha' \sqrt{2T_3}}
\left( 1-\frac{v^2}{4} \right)$.
In fact, the Yang-Mills theory, similar to
the effective action of the brane, includes
information about the brane. Therefore,
various properties of the branes can be reliably described
in the language of the Yang-Mills theory \cite{37}.

\subsubsection{The brane cosmology}

Here we give a brief speculation around the brane cosmology
corresponding to a dynamical brane.
For example, we consider the action (3.2).
Quench the $B$-field and the
$U(1)$ gauge potential. By introducing a DBI
field $\phi$ and an appropriate potential $V(\phi)$
we receive the effective action
\begin{eqnarray}
I=&-&\int d^4 x \bigg[ \frac{1}{f(\phi)}\sqrt{-\det
\left( {\tilde G}_{\alpha \beta}
+8\pi\alpha'\omega_{\alpha \beta}
+f(\phi)\partial_\alpha \phi\partial_\beta \phi
\right)}
\nonumber\\
&+&\sqrt{-\det{\tilde G}_{\alpha \beta}}\left(-\frac{1}{f(\phi)}
+V(\phi)\right)\bigg],
\end{eqnarray}
where the free functional $f(\phi)$ is related to the inverse
of the  D3-brane tension, and is specified by fixing the
cosmological model, e.g. see \cite{38, 39}.
The reparametrization symmetry induces a gauge freedom,
which was fixed by selecting the static gauge
$\{x^\alpha=\xi^\alpha|\alpha=0,1,2,3\}$.
For the D3-brane with the low-energy tangential
dynamics this action takes the from
\begin{eqnarray}
I'=&-&\int d^4 x \sqrt{-\det{\tilde G}_{\alpha \beta}}
\bigg[ \frac{1}{f(\phi)}\sqrt{1+
32(\pi\alpha')^2\omega^{\alpha \beta}\omega_{\alpha \beta}
+f(\phi){\tilde G}^{\alpha \beta}\partial_\alpha
\phi\partial_\beta \phi }
\nonumber\\
&-&\frac{1}{f(\phi)}+V(\phi)\bigg].
\end{eqnarray}
For a stationary brane
this action reduces to the conventional action
of the literature, e.g. see \cite{38}.
By ignoring the spatial derivatives of
$\phi$, i.e. for an approximately
homogeneous DBI field,
the third term under the square root finds the
feature $-f(\phi){\dot \phi}^2$.

Now let us define the Lorentz-like factor
\begin{eqnarray}
\Gamma_\omega=1/\sqrt{1+
32(\pi\alpha')^2\omega^{\alpha \beta}\omega_{\alpha \beta}
-f(\phi){\dot \phi}^2}~.
\end{eqnarray}
For a constant DBI field and in the absence of the tangential
rotation, i.e. for $\omega_{{\bar \alpha}{\bar \beta}}=0$
with ${\bar \alpha},{\bar \beta}\in\{1,2,\cdots,p\}$,
this Lorentz-like factor manifestly reduces to the usual
Lorentz factor of special relativity with the velocity
components
$V_{\bar \alpha}=4\pi\alpha'\sqrt{2}\omega_{0{\bar \alpha}}$.
We observe that the scalar
field cannot roll down arbitrarily fast. Its
rolling is controlled by the positivity of
the phrase under the square root in Eq. (3.10),
the linear velocity
$\omega_{0{\bar \alpha}}$ and the angular
velocity $\omega_{{\bar \alpha}{\bar \beta}}$ of the brane.

According to the action (3.9) the energy
density and pressure of the DBI field possess the forms
\begin{eqnarray}
&~&\rho_\phi=\rho^{(0)}_\phi
+(\Gamma_\omega-\Gamma_0)\frac{1}{f(\phi)},
\nonumber\\
&~&p_\phi=p^{(0)}_\phi
+\left(\frac{1}{\Gamma_0}-\frac{1}{\Gamma_\omega}\right)
\frac{1}{f(\phi)},
\end{eqnarray}
where $\rho^{(0)}_\phi$
and $p^{(0)}_\phi$ exhibit the energy density
and pressure of the DBI field, associated with the
stationary brane. It is usually assumed that
$f(\phi)$ to be non-negative. Therefore, since there is
$\Gamma_\omega < \Gamma_0$, we acquire
$\rho_\phi<\rho^{(0)}_\phi$ and $p_\phi<p^{(0)}_\phi$.
In the same way, the modification of the energy-momentum
tensor is given by
\begin{eqnarray}
T_{\alpha \beta}=T^{(0)}_{\alpha \beta}
+\left(\frac{1}{\Gamma_\omega}-\frac{1}{\Gamma_0}\right)
\frac{1}{f(\phi)}{\tilde G}_{\alpha \beta}.
\end{eqnarray}
Eqs. (3.11) and (3.12) imply that the brane dynamics
extremely modifies the main quantities of the brane cosmology.
However, by applying the action (3.9) and Eqs. (3.11) and
(3.12) one may perform the principal equations to
investigate the behavior of the corresponding brane
cosmology. For example, the inflation and dark energy
solutions, extracted from  the DBI models \cite{40},
will be obviously improved by the brane dynamics.

\section{The Wess-Zumino like action}

It is known that a D-brane
carries an R-R charge \cite{1}. This implies that
there are couplings between the massless R-R fields
and the brane.
The effective action which accurately
specifies these interactions
is the Wess-Zumino action \cite{3},
\cite{19}-\cite{21}. The corresponding
Lagrangian can be naturally obtained by computing the
inner product between the states $|C_n \rangle$,
representing the massless R-R
states, and the boundary state of the R-R sector
\cite{3}, i.e.,
\begin{eqnarray}
\mathcal{L}_{\rm WZ} \varpropto
\langle{C_n}| B \rangle_{\rm R}~,
\end{eqnarray}
where $n$ is odd (even) for the type IIA (type IIB) theory.
The massless R-R states $|C_n \rangle$,
in the picture $(-1/2,-3/2)$, can be expressed as \cite{3},
\begin{eqnarray}
|C_n\rangle &=&
\frac{1}{2 \sqrt{2} n!} C_{\mu_{1}\ldots\mu_{n}}\bigg{[}
\left({C}\Gamma^{\mu_{1}\ldots\mu_{n}}
\Pi_{+}\right)_{AB}~\cos (\gamma_0\tilde{\beta_0})
\nonumber\\
&+&\left({C}\Gamma^{\mu_{1}\ldots\mu_{n}}
\Pi_{-}\right)_{AB}~\sin (\gamma_0\tilde{\beta_0})
\bigg{]}~|A; k/2\rangle_{-1/2} \otimes|\tilde{B};
\widetilde{k/2}\rangle_{-3/2}~,
\end{eqnarray}
where $\Gamma^{\mu_{1}\ldots\mu_{n}}$ is the antisymmetrized
product of the matrices
$\{\Gamma^{\mu_1}, \Gamma^{\mu_2}, \ldots, \Gamma^{\mu_n}\}$,
$\beta_{0}$ and $\gamma_{0}$
are the superghost zero-modes, and
$\Pi_{\pm}=(1\pm \Gamma_{11})/2$.
The state $|C_n\rangle$ is directly associated with
the $n$-form R-R potential $C_n$.

Now we can compute a Wess-Zumino like action for
the dressed-dynamical brane.
The coupling between the brane and the
R-R potential $C_n$ is explicitly given by the
overlap between the states (4.2) and the boundary state
of the R-R sector
\begin{eqnarray}
\langle{C_n}| B \rangle_{\rm R}&=&-\frac{T_{p}}{16 \sqrt{2} n!}
\frac{V_p}{v}
~ C_{\mu_{1}\ldots \mu_{n}} {\rm Tr}
\left(\Gamma^{\mu_{n}\ldots\mu_{1}}
\left( \Gamma^{0} + v\Gamma^{i_0}\right)\Gamma^1 \ldots
\Gamma^{p}~\mathcal{H}
:e^{-\frac{1}{2}{\Phi}_{\lambda\lambda'}
\Gamma^{\lambda}\Gamma^{\lambda'}}:\right),
\nonumber\\
\mathcal{H} &=& \left[ {\bf 1}+v\Gamma^{i_0}\Gamma^{0}
-2v\Gamma^{{i_0}}\Gamma^{0}
\left(1+(PQ^{-1}N)^{i_0}_{~~\lambda}\Gamma^{i_0}
\Gamma^{\lambda}\right)^{-1}\right]^{-1}.
\end{eqnarray}
Calculation of the above trace for an arbitrary
velocity is very complicated. For simplification
we assume that the brane moves with a small velocity,
i.e. $v<<1$. Furthermore, we consider the indices
$\mu_1, \ldots, \mu_n$ along the worldvolume
of the brane. Consequently, the exponential part
and the matrix $\mathcal{H}$ reduce to
\begin{eqnarray}
:e^{-\frac{1}{2}{{\Phi}_{{\lambda \lambda'}}
\Gamma^{\lambda}\Gamma^{\lambda'}}}: &=& \left[
1-\frac{v}{2}
\left(\frac{\partial {\Phi}_{\lambda \lambda'}}
{\partial v}\right)_{v=0}~
:\Gamma^{\lambda}\Gamma^{\lambda'}:+\mathcal{O}(v^2)\right]
:e^{-\frac{1}{2}{{\Phi}^{(0)}_{{\lambda_1 \lambda_2}}
\Gamma^{\lambda_1}\Gamma^{\lambda_2}}}:,
\nonumber\\
\mathcal{H} &=& {\bf 1}+v\Gamma^{{i_0}}\Gamma^{0}
-2v (PQ_{(0)}^{-1}{N_{(0)}})^{i_0}_{~~\lambda}\Gamma^{0}
\Gamma^{\lambda}+\mathcal{O}(v^2),
\end{eqnarray}
where ${\Phi}^{(0)}_{\lambda_1 \lambda_2}$,
$Q_{(0)}$ and $N_{(0)}$ are
${\Phi}_{\lambda_1 \lambda_2}$,
$Q$ and $N$ with $v=0$, respectively.
Now by expanding the exponential factor
of the right-hand side of Eq. (4.4),
due to the antisymmetrization symbol
$:~:$, different exponents of ${\Phi}^{(0)}$
of the rank $l\in \{0,1,\ldots,l_{\rm max}\}$
will appear in the first equation of (4.3).
For receiving a nonzero trace
we determine $l_{\rm max}$ via the velocity
independent term inside the trace part. Therefore,
we acquire $l_{\rm max}=p/2$ ($l_{\rm max}=(p+1)/2$)
for the type IIA theory (type IIB theory),
and we should use $n=p+1-2l$.

At first we apply $n=p+1$, which
indicates the coupling of the D$p$-brane with
the $(p + 1)$-form potential $C_{p+1}$,
\begin{eqnarray}
\langle{C_{p+1}}| B \rangle_{\rm R}
&=&\frac{\sqrt{2} T_{p}}{(p+1)!}
~ \frac{V_p}{v}{\tilde C}_{\alpha_0 \ldots \alpha_p}~
\varepsilon^{\alpha_{0}\ldots\alpha_{p}}
\bigg[ 1+2v (PQ_{(0)}^{-1}N_{(0)})^{i_0}_{~~0}
\nonumber\\
&+& 2v (PQ_{(0)}^{-1}N_{(0)})^{i_0}_{~~\bar{\lambda}}
{{\Phi}}^{(0)}_{{0\bar{\lambda}}}
-\frac{v}{4}\left( \frac{\partial {{\Phi}}_{\alpha \beta}}
{\partial v}\right)_{v=0}~{{\Phi}}^{(0)}_{\alpha \beta}
+\mathcal{O}(v^2)~\bigg],
\end{eqnarray}
where $\bar{\lambda}\in \{1,2,\ldots,p,i_0\}$,
$V_p$ is the brane volume,
$\varepsilon^{\alpha_{0} \ldots \alpha_{p}}$ is the
Levi-Civita tensor, and the tensor
${\tilde C}_{\alpha_0 \ldots \alpha_p}$ is the
pull-back of $C_{\mu_0 \ldots \mu_p}$ on the
brane worldvolume.

The next case is $n=p-1$, which
defines the following interaction terms
\begin{eqnarray}
\langle{C_{p-1}}| B \rangle_{\rm R}
&=&-\frac{\sqrt{2}T_{p}}{2(p-1)!}
~\frac{V_p}{v} \bigg{\{}
{\tilde C}_{\alpha_{0} \ldots \alpha_{p-2}}
\varepsilon^{\alpha_{0} \ldots \alpha_{p}} \bigg{[}
{\Phi}^{(0)}_{\alpha_{p-1}\alpha_{p}}
\bigg{(1}+2v (PQ_{(0)}^{-1}N_{(0)})^{i_0}_{~~0}
\nonumber\\
&+& 2v(PQ_{(0)}^{-1}N_{(0)})^{i_0}_{~~\bar{\lambda}}
{{\Phi}}^{(0)}_{{0\bar{\lambda}}}
-\frac{v}{4}
\left( \frac{\partial {{\Phi}}_{\alpha \beta}}
{\partial v}\right)_{v=0}~
 {{\Phi}}^{(0)}_{\alpha \beta}\bigg{)}
\nonumber\\
&-&\frac{v}{4}
{\Phi}^{(0)}_{i_0\alpha_{p-1}}
\left( \frac{\partial {{\Phi}}_{i_0\alpha_{p}}}
{\partial v}\right)_{v=0}
+v \left( \frac{\partial {{\Phi}}_{{\alpha}_{p-1}
{\alpha}_{p}}}{\partial v}\right)_{v=0}
~ \bigg{]}
\nonumber\\
&+& 4 v{\tilde C}_{{\bar{\alpha}}_{1} \ldots {\bar{\alpha}}_{p-1}}
\varepsilon^{{\bar{\alpha}}_{1} \ldots {\bar{\alpha}}_{p-1}
{\bar{\alpha}}_{p}}
(PQ_{(0)}^{-1}N_{(0)})^{i_0}_{~~{\bar{\alpha}}_{p}}
\bigg{\}}.
\end{eqnarray}
The first R-R interaction
represents the coupling of the potential $C_{p-1}$
with the D$p$-brane. This coupling clearly comprises
all components of the pull-back tensor
${\tilde C}_{\alpha_{0} \ldots \alpha_{p-2}}$.
The second R-R interaction reveals the
coupling of the D$p$-brane with the
same potential $C_{p-1}$.
This coupling includes only
the pure spatial components of
${\tilde C}_{\alpha_0 \ldots \alpha_{p-2}}$, i.e.
${\bar{\alpha}}_{1}, \ldots , {\bar{\alpha}}_{p-1} \neq 0$.

The coupling of the brane with the potential $C_{p-3}$
is given by
\begin{eqnarray}
\langle{C_{p-3}}| B \rangle_{\rm R}
&=&\frac{\sqrt{2}T_{p}}{ 4(p-3)!}
~ \frac{V_p}{v} \bigg{\{}
{\tilde C}_{\alpha_{0}\ldots\alpha_{p-4}}
\varepsilon^{\alpha_{0}\ldots\alpha_{p}}
{\Phi}^{(0)}_{\alpha_{p-3}\alpha_{p-2}}
\bigg{[}{\Phi}^{(0)}_{\alpha_{p-1}\alpha_{p}}
\bigg{(}1+2v(PQ_{(0)}^{-1}N_{(0)})^{i_0}_{~~0}
\nonumber\\
&+& 2v (PQ_{(0)}^{-1}N_{(0)})^{i_0}_{~~\bar{\lambda}}
{{\Phi}}^{(0)}_{{0\bar{\lambda}}}
-\frac{v}{4}
\left(\frac{\partial {{\Phi}}_{\alpha \beta}}
{\partial v}\right)_{v=0}~
{{\Phi}}^{(0)}_{\alpha \beta}\bigg{)}
\nonumber\\
&-&~\frac{v}{2}
{\Phi}^{(0)}_{i_0\alpha_{p-1}}
\left( \frac{\partial {{\Phi}}_{i_0\alpha_{p}}}
{\partial v}\right)_{v=0} + v
\left( \frac{\partial {{\Phi}}_{{\alpha}_{p-1}
{\alpha}_{p}}}{\partial v}\right)_{v=0}
~\bigg{]}
\nonumber\\
&+& 4v {\tilde C}_{{\bar{\alpha}}_{1}\ldots{\bar{\alpha}}_{p-3}}
\varepsilon^{{\bar{\alpha}}_{1}\ldots
{\bar{\alpha}}_{p}}{\Phi}^{(0)}_{{\bar{\alpha}}_{p-2}
{\bar{\alpha}}_{p-1}}
(PQ_{(0)}^{-1}N_{(0)})^{i_0}_{~~{\bar{\alpha}}_{p}}\bigg{\}},
\end{eqnarray}
The first R-R interaction
shows the coupling of the potential $C_{p-3}$
with the D$p$-brane, which contains
all components of the pull-back tensor
${\tilde C}_{\alpha_{0} \ldots \alpha_{p-4}}$.
The second R-R interaction clarifies the
coupling of the brane with the same potential $C_{p-3}$.
This coupling only consists of
the pure spatial components of
${\tilde C}_{\alpha_0 \ldots \alpha_{p-4}}$.

In the  same way one can obtain couplings of the brane
with the other R-R potentials $C_n $ for $n \leq p-5$.
These couplings, accompanied by Eqs. (4.5)-(4.7), establish
the following Wess-Zumino like action
\begin{eqnarray}
S^{(\omega , v)}_{\rm WZ}&=& \frac{\mu_p V_p}{vV_{p+1}}
\int_{V_{p+1}}\bigg{\{}\left[
\sum_{l=0}^{l_{\rm max}} C_{p+1-2l}\wedge
e^{2\pi\alpha'{\Phi}^{(0)}}\right]_{p+1}
\nonumber\\
&+& vC_{p+1}\bigg[ 2(PQ^{-1}N)^{i_0}_{~~0}
+ 2(PQ^{-1}N)^{i_0}_{~~\bar{\lambda}}
{{\Phi}}_{{0\bar{\lambda}}}
-\frac{1}{4}\frac{\partial {{\Phi}}_{\alpha \beta}}
{\partial v}{{\Phi}}_{\alpha \beta}\bigg]_{v=0}
\nonumber\\
&-& 2\pi\alpha' v C_{p-1}\wedge \bigg{[}{{\Phi}}
\bigg{(}2 (PQ^{-1}N)^{i_0}_{~~0}
+ 2(PQ^{-1}N)^{i_0}_{~~\bar{\lambda}}
{{\Phi}}_{{0\bar{\lambda}}}
\nonumber\\
&-& \frac{1}{4}
\frac{\partial {{\Phi}}_{\alpha \beta}}
{\partial v}{{\Phi}}_{\alpha \beta}\bigg{)}
- \frac{1}{4} \Omega
+ \frac{\partial{{\Phi}}}{\partial v}\bigg{]}_{v=0}
+\frac{1}{2}(2\pi\alpha')^2v C_{p-3}\wedge {{\Phi}}^{(0)}
\nonumber\\
&\wedge & \bigg{[}{{\Phi}}
\bigg{(}2(PQ^{-1}N)^{i_0}_{~~0}
-\frac{1}{4}
\frac{\partial {{\Phi}}_{\alpha \beta}}{\partial v}
{{\Phi}}_{\alpha \beta}
+  2 (PQ^{-1}N)^{i_0}_{~~\bar{\lambda}}
{{\Phi}}_{{0\bar{\lambda}}}\bigg{)}
\nonumber\\
&-& \frac{1}{2}\Omega
+ \frac{\partial{{\Phi}}}{\partial v}
\bigg{]}_{v=0}+\ldots \bigg{\}}
\nonumber\\
&-& 8\pi\alpha'\mu_p v\int_{V_p}\left[ \left(
{\bar C}_{p-1} - \frac{1}{2}\pi\alpha'
{\bar C}_{p-3}\wedge {\bar \Phi}^{(0)}
\right) \wedge {\bar W}+ \ldots\right] ,
\end{eqnarray}
where $\mu_p =\sqrt{2}T_p$ is the R-R charge
of the brane because of the potential $C_{p+1}$.
The differential forms ${\Phi}$, ${\bar \Phi}^{(0)}$, $\Omega$
and ${\bar W}$ have the following definitions
\begin{eqnarray}
{\Phi}&=& \frac{1}{2}{{\Phi}}_{\alpha \beta}~
d\xi^\alpha \wedge d\xi^\beta,
\nonumber\\
{\bar \Phi}^{(0)}&=& \frac{1}{2}
{\Phi}^{(0)}_{{\bar \alpha}{\bar \beta}}~
d\xi^{\bar \alpha} \wedge d\xi^{\bar \beta},
\nonumber\\
\Omega &=&
\frac{1}{2}\left( {{\Phi}}_{i_0 \alpha}
\frac{\partial{{\Phi}}_{i_0 \beta}}{\partial v}
-{{\Phi}}_{i_0 \beta}
\frac{\partial{{\Phi}}_{i_0 \alpha}}{\partial v}
\right)_{v=0} d\xi^\alpha \wedge d\xi^\beta ,
\nonumber\\
{\bar W}&=&
( PQ^{-1}_{(0)}N_{(0)})^{i_0}_{~~{\bar \alpha}}
d\xi^{\bar \alpha}~.
\end{eqnarray}
The R-R forms in the last integral
possess only the pure spatial components, and are defined by
\begin{eqnarray}
{\bar C}_m=\frac{1}{m!}{\tilde C}_{{\bar \alpha}_1
{\bar \alpha}_2 \ldots {\bar \alpha}_m}
d\xi^{{\bar \alpha}_1}\wedge
d\xi^{{\bar \alpha}_2} \wedge\ldots \wedge d\xi^{{\bar \alpha}_m} ,
\end{eqnarray}
where $m \in \{p-1, p-3, p-5, \ldots \}$.

Note that the matrices $Q,~Q_{(0)}$, $N,~N_{(0)}$,
${{\Phi}}$ and ${{\Phi}}^{(0)}$
explicitly depend on the potential
$A_\alpha (\xi^0 ,\xi^1 ,\ldots , \xi^p)$, via its
field strength $F_{\alpha \beta}$,
and the worldvolume coordinates
$X^\mu(\xi^0 ,\xi^1 ,\ldots , \xi^p)$.
Since the gauge field and worldvolume coordinates
are the main degrees of freedom,
the WZ-like and the DBI-like actions
exhibit a generalized effective action for
the dressed-dynamical brane.
However, the effective Lagrangian
is a very complicated functional of
the foregoing degrees of freedom.

Finally, by stopping the brane, i.e. by setting
$\omega$ and $v$ to zero, we receive
$\Phi_{\alpha\beta}=\mathcal{F}_{\alpha\beta}$,
and hence the WZ-like action (4.8) reduces to the
conventional WZ action, as expected.

We extended the effective action of a D$p$-brane
via its tangential and transverse dynamics.
In fact, there are various extensions for the
brane effective action: derivative corrections \cite{33},
$\alpha'$-corrections \cite{41}, tachyonic extension
\cite{42} and curvature corrections \cite{43}.
Accordingly, for a given setup of a D$p$-brane
one may combine some of these modifications to construct a
suitable action. For example, for
a dynamical D$p$-brane with the tachyon field in a curved
background one should add an appropriate tachyon potential
and the curvature improvements to our action.

\section{Conclusions}

We obtained the effective action of a
dynamical D$p$-brane with background fields.
This generalized action consists of the DBI-like and WZ-like
parts. To obtain the effective action we applied
the boundary state formalism with the
following background fields: a constant Kalb-Ramond field and
a $U(1)$ internal gauge potential.
The dynamics of the brane includes a tangential rotation
and a linear motion with both tangential and
transverse components. For slow motion of the brane
we decomposed the DBI-like action into a pure DBI one
and its corrections due to the brane dynamics.
We acquired the WZ-like action by computing
the couplings of the R-R sector boundary state
with the massless states of the R-R sector.

The effective action of the brane
depends on the brane velocities $v$ and
$\omega_{\alpha\beta}$, and the
fields $F_{\alpha\beta}$ and $B_{\mu\nu}$.
The variety of the variables
$\{F_{\alpha\beta}, B_{\mu\nu},\omega_{\alpha\beta},v;p\}$
dedicated a generalized feature to the action. However,
by expanding the determinant and then
square root in the DBI-like part, and also
$Q^{-1}$ and exponential in the WZ-like part, one can
read the coupling constants. These constants
depend on the input parameters $\omega_{\alpha\beta}$ and $v$.
Thus, by adjusting the values of these parameters,
the values of the coupling constants
can be accurately adjusted to any desirable values.


\end{document}